# Summary Report of Working Group 5: Beam and Radiation Generation, Monitoring, and Control

Mike Church[*] and Kiyong Kim[†]

[*]Fermi National Accelerator Lab, PO Box 500, Batavia, IL 60510
[†]Department of Physics, University of Maryland, College Park, MD 20742-4111

**Abstract.** This paper summarizes the activities and presentations of Working Group 5 of the Advanced Accelerator Concepts Workshop held at Annapolis, Maryland in June 2010. Working Group 5 touched on a broad range of topics in the fields of beam and radiation generation and their monitoring and control. These topics were not comprehensively covered in this Workshop, but rather the Working Group concentrated on specific new developments and recent investigations. The Working Group divided its sessions into four broad categories: cathodes and electron guns, radiation generation, beam diagnostics, and beam control and dynamics. This summary is divided into the same structure.

## WORKING GROUP FOCUS AND STRUCTURE

Our focus was on research that is relatively new (published over the last few years), and has promise to advance the state-of-the-art in electron beam sources, electron beam control and manipulation, radiation generation (THz to Gamma-ray), and electron beam diagnostics, both intercepting and non-intercepting.

Presentations on electron sources covered new gun technology, including DC guns, RF guns, and development of a superconducting RF gun. New results on both photocathodes and thermionic cathodes were presented along with the latest progress on the development of a practical diamond-amplified photocathode.

Discussions on beam control concentrated on methods of partitioning emittances to optimize beam parameters for particular applications. The performance, latest results, and upgrade plans of the low energy electron ring at the University of Maryland (UMER) were presented in detail. The concept of adiabatic thermal beams was proposed by Chiping Chen (MIT) as a new method to control beam halo and beam loss.

Radiation generation presentations focused on new methods of generating terahertz to fill the well-known "terahertz gap" in the EM spectrum. Five very distinct approaches were discussed, ranging from high-power gyrotrons to free electron lasers (IFELs). New results in X-ray generation from laser wakefield accelerators was discussed in the joint Working Group 1/4/5 session (see Working Group 1 summary for details). X-ray generation from inverse Compton scattering at the T-REX facility at LLNL were presented by Felicie Albert (LLNL) as well as plans to upgrade that facility to generate Gamma-rays.

Electron beam diagnostics is a broad area of research and a selection of promising techniques were presented in this Working Group. The current challenges in optical transition radiation (OTR) techniques were reviewed by Ralph Fiorito (UMD) and the current plans for advancing optical diffraction radiation (ODR) non-intercepting techniques were reviewed by Alex Lumpkin (FNAL). Several promising methods of using terahertz radiation from electro-optic sampling were presented. Finally, an interesting new technique to image beam halo was presented by Hao Zhang (UMD).

Working Group 5 met for 10 different sessions over 5 days. These sessions were grouped into 4 distinct categories: cathodes and electron guns (2 sessions); radiation generation (2 sessions + 1 joint session with Working Groups 1 and 4); beam diagnostics (2 sessions); and beam control and dynamics (3 sessions). There were a total of 37 oral presentations, 15 poster presentations, and 2 plenary presentations. There was also very strong student participation, with 12 students making presentations (oral and/or poster). Table 1 lists all the contributions by category.

TABLE 1. Categorized presentations for Working Group 5.

| Category | Subject | Authors |
|---|---|---|
| Cathodes | Diamond amplified | I. Ben-Zvi |
| | Photocathode | P. Musumeci, K. Nemeth, M. Uesaka |
| | Thermionic | L. Ives |
| Electron guns | SRF | J. Lewellen |
| | DC | J. Zhou |
| | RF | C. Neumann |
| THz generation | Two-stream instability | K. Bishofberger |
| | Smith-Purcell | P. Piot |
| | Gyrotron | M. Glyavin |
| | IFEL | S. Tochitsky |
| | Corrugated plasma | A. Pearson |
| X-ray generation | Compton scattering | T. Natsui, F. Albert |
| | Betatron radiation | S. Kneip |
| | Laser-driven undulator | F. Gruner |
| γ-ray generation | Compton scattering | F. Albert |
| Intercepting beam diagnostics | OTR | R. Fiorito, A. Lumpkin |
| | Halo imaging | H. Zhang, R. Fiorito |
| Non-intercepting beam diagnostics | ODR | A. Lumpkin |
| | THz EOS | C. Scoby, M. Helle, J. van Tilborg |
| | CSR | J. Thangaraj |
| Beam control | Emittance exchange | P. Piot, J. Power, B. Carlston, D. Xiang, J. Ruan |
| | CSR Suppression | M. Fedurin |
| | Beam compression with terahertz | J. Moody |
| Beam dynamics | UMER | R. Kishek, S. Bernal, B. Beaudoin, T. Koeth, K. Fiuza |
| | Adiabatic thermal beams | C. Chen |

# CATHODES AND ELECTRON GUNS

The accelerator community currently has a strong demand for higher brightness ultrafast electron sources, and several avenues towards this goal were presented at this Workshop. I. Ben-Zvi gave an update on progress in the development of a diamond-amplified photocathode for use in an electron gun. A diamond crystal is placed directly behind a photocathode with a HV potential between them, and the accelerated photoelectrons are amplified in the diamond by secondary emission and emitted from the hydrogenated opposite side of the diamond crystal. Average currents of 15A/cm$^2$ have been measured with electron gain of >40, temporal spread of a few ps, and excellent diamond lifetime. The measurements agree very well with simulations done by Tech-X Corporation, and the simulations indicate a thermal emittance of <0.05 μm. The diamond amplifier serves to protect the gun from the photocathode, and vice-versa, and therefore has obvious potential application in SRF guns.

Operating photocathodes in the "blow-out" regime wherein the space charge forces are linear was first proposed by Serafini, et al. in 1997 [1]. The first demonstration of this principle was made from metal photocathodes in 2008 by Musumeci, et al. [2]. In this Workshop P. Musumeci presented the latest results from the Pegasus Laboratory at UCLA on this method of generating low emittance, ultrashort electron bunches in the linear space charge regime using a Ti:Sa laser with pulse length 35 fs to excite the photocathode. At 20 pC bunch charge, an emittance of 0.5 μm and rms bunch length of 300 fs is obtained. At the Pegasus Laboratory pump-and-probe femtosecond relativistic electron diffraction experiments to study ultrafast phase transitions are being done with these beams, which have 3 orders of magnitude higher intensity than conventional electron diffraction sources. In addition, investigations of multiphoton (IR) induced electron emission from metal cathodes have found surprisingly high quantum efficiencies. This technique has the advantage that it foregoes the inefficient and often troublesome step of doubling the laser IR frequency to produce UV photons.

Further investigations of photocathodes were described by K. Nemethy and M. Uesaka. Nemethy has been performing theoretical calculations of oxide layers on metal nanostructures with the aim of producing a potentially higher brightness photocathode. He proposed a practical design consisting of 50 – 200 nm Ag rods in an MgO substrate with ultrathin layers of MgO to reduce the work function and with possibility of thermal emittance as low as 0.06 µm. Uesaka described ongoing developments at the Femtosecond Electron Linac at the University of Tokyo, including improvements to their $Na_2KSb$ photocathode and some plans to develop an antimonic cathode.

A recent development in thermionic cathodes was described by L. Ives of Calabazas Creek Research. They have developed a cathode consisting of an emitter of sintered tungsten wire with ~4 µm pores with 10 – 20 µm spacing with a barium reservoir situated behind the emitter. The barium diffuses through the pores as the cathode is heated. This design has led to a dramatic improvement of cathode lifetime over traditional scandate cathodes -- a lifetime of 32,000 hours at 50 A/$cm^2$ has been measured.

New electron gun designs were described by J. Lewellen and J. Zhou. Lewellen showed the $1^{st}$ beam results from the prototype NPS SRF gun designed and built by Niowave. This SRF gun is a 500 MHz ¼ wave resonator with a 100 W RF source, metal cathode, and superconducting solenoid for emittance compensation. It's initial purpose will be to provide a source for nC-range bunch charge dynamics studies for a future compact ERL, but it also has the obvious possibility of testing novel photocathodes, such as field emitters. Eventually this design will be the injector for a 10 mA NPS FEL. J. Zhou (Beam Power Technology) presented beam imaging results from a prototype elliptic beam gun, intended for application in L-band elliptic beam klystrons and possibly for beam experiments with planar structures which require high aspect ratio beams. This gun operates at 2.3 kV, 110 mA, and a 6:1 aspect ratio. The beam imaging measurements are in good agreement with simulations (OMNITRAK).

## RADIATION GENERATION

In the THz regime, there was an interest in using low energy (<100 keV) electron beams for THz generation. This includes two-stream instability, Smith-Purcell, and gyrotrons as shown in Table 1. First, K. Bishofberger discussed particle-in-cell (PIC) simulations of THz generation using two-stream instability. This scheme is not new but it uses two-stream instability as a gain mechanism in a klystron framework. Upon mixing two electron beams (for example, 20 keV and 19.5 keV), beam bunching can occur via the instability and saturate within tens of centimeters. Finally, a portion of the bunched beam energy can be extracted and radiated at THz frequencies by coherent synchrotron radiation (CSR) or coherent transition radiation (CTR). Simulations show significant THz radiation (up to several Watts) at 0.03 ~ 1 THz.

P. Piot discussed a two-stage Smith-Purcell THz radiation source. Compared to a conventional single-stage Smith-Purcell source, in which beam bunching and THz radiation occur on the same grating, the two-stage method separates those two steps. The first stage (buncher) consisting of a pair of gratings can enhance beam modulation by a factor of 2 and provide better beam quality compared to a single-stage source [3]. The second stage (radiator) consisting of a single grating can emit THz radiation. The expected average power is of the order of Watts. The radiation frequency is tunable by varying the electron beam energy or altering the first stage grating gap. This novel scheme was confirmed with 2D finite-difference time-domain (FDTD) VORPAL simulations [3], and its experimental demonstration is under development.

Gyrotrons are another device using low energy beams for THz generation. M. Glyavin presented his recent development of powerful THz gyrotrons capable of producing several kW power [4]. With a 50 T pulsed magnet, the output power has reached 5 kW (210 mJ over 40 µs) at 1 THz and 0.5 kW (10 mJ over 20 µs) at 1.3 THz. However, the repetition rate is limited to one shot per minute because of significant heating in the solenoid. Currently, Glyavin and his group are developing high-repetition-rate and CW gyrotrons, as well as more powerful (1MW) sub-THz gyrotrons.

High-power THz radiation can be also obtained in free electron lasers (FELs). S. Tochitsky at the UCLA Neptune laboratory discussed the development of a single-pass FEL capable of producing multi-MW (peak power) radiation tunable in the range of 0.5-9 THz. The Neptune group has ready demonstrated a THz seed source producing ~1 kW peak power (0.4 mJ over 200 ns) at 0.5-3 THz via difference frequency mixing of $CO_2$ lasers in GaAs. Tochitsky, et al. also demonstrated microbunching of 12 MeV electron beams in high-order inverse free electron laser (IFEL) interactions [5]. These microbunched beams can be further used for a matched injection in laser-plasma accelerators or for producing THz radiation via CSR or CTR.

Intense THz radiation can be also produced from laser pulses interacting with plasma. A. Pearson at UMD discussed his simulation results on THz generation by laser pulses propagating in corrugated plasma channels, where the electron density is longitudinally modulated [6]. In this scheme, the ponderomotive force of laser pulses excites THz radiation in the plasma channel. Yet, for optimal conversion from laser to THz radiation, the phase velocity of radiation modes should be matched with the group velocity of the current source. This is exactly what happens in the corrugated plasma channel. Simulations show that ten percent of laser energy may be converted to THz radiation with a realistic electron density profile.

X-ray and gamma-ray generation via Compton scattering was discussed by T. Natsui and F. Albert. Compton scattering occurs when laser pulses are scattered by a relativistic electron beam, which generates X-ray and gamma-ray pulses. Natsui presented the development of an X-band linac system for Compton scattering X-ray sources at University of Tokyo. Albert presented the status of the Thomson Radiated Extreme X-ray (T-REX) Compton scattering source at LLNL, and its application for nuclear resonance fluorescence detection [7]. She also presented the design of a future Compton scattering source capable of producing monoenergetic, tunable gamma-rays in the 0.5-2.5 MeV range.

Other novel schemes for X-ray generation, including betatron radiation in laser wakefield acceleration and laser-driven undulator radiation sources, were discussed in the joint session and are described in the WG 1 Summary.

## ELECTRON BEAM DIAGNOSTICS

Beam diagnostics are of great importance in characterizing beam size, bunch-length, divergence, and emittance. The diagnostics discussed in WG 5 include coherent and incoherent optical transition radiation (OTR), optical diffraction radiation (ODR), halo imaging, and electro-optic (EO) THz diagnostics (see Table 1).

Transition radiation (TR) occurs when a charged bunch traverses a metallic foil, and this radiation provides useful information on the beam size and bunch length. R. Fiorito reviewed the state of the art and challenges in optical TR (OTR) beam imaging diagnostics. In particular, coherent TR (CTR) occurs when the bunch length is shorter than the radiation wavelength, and can be used for monitoring and/or measuring the bunch length with an additional setup of autocorrelation interferometry. OTR has been also used for imaging low energy (10 keV) beams at the University of Maryland Electron Ring (UMER). H. Zhang at UMER described a new method for imaging beam halos with an adaptive optical mask. In this scheme, a digital micromirror array was used to deflect the intense inner core light beam, which otherwise could saturate and possibly damage the camera. This technique allows high dynamic range ($>10^5$) imaging of faint halo structures, useful for understanding beam dynamics and control.

Optical diffraction radiation (ODR) diagnostics, similar to OTR but non-intercepting beam size monitors, were reviewed by A. Lumpkin. Diffraction radiation (DR) occurs when a charged beam passes nearby a boundary between media having different dielectric constants. For highly relativistic beams, a metal slit at a distance larger than the transverse beam size (thus non-intercepting) can produce DR at optical frequencies. Lumpkin described ODR generation by a metal plane with a slit aperture, a single metal plane, and two-foil interferometer. He also presented the results of ODR far-field imaging at KEK and FLASH, as well as near-field results at Advanced Photon Source (APS) facility at Argonne, Continuous Electron Beam Accelerator Facility (CEBAF) at JLab, and FLASH. Lumpkin also discussed a promising path towards testing near-field imaging on 10 μm size, 25 GeV beams at FACET.

Electro-optic (EO) based THz diagnostics are of great current interest as a non-invasive method, in particular for characterizing ultrashort electron bunches. In this scheme, the Coulomb field arising from the electron bunch, which represents the transverse and longitudinal bunch profiles, induces transient birefringence in an EO crystal located near the beam (thus non-intercepting). Ultrashort laser pulses synchronized with the beam can be used to measure the birefringence, which directly maps out the electric field (or bunch profile). The current interest is to extend the technique to characterizing ultrashort electron beams (<10 fs) in laser wake field acceleration and X-ray free electron lasers. In this effort, M. Helle at NRL analyzed some potential limitations in THz EO detection where the crystal dispersion and absorption (linear and nonlinear) can greatly distort the THz temporal waveform and thus the bunch profile. In addition, Helle proposed an improved EO diagnostic for ultrashort electron bunch measurements using X-FROG, which holds great promise for characterizing e-beams produced by laser wakefield accelerators.

C. Scoby at the UCLA Pegasus photoinjector laboratory discussed a spatially encoded EO technique and its use as a time-of-arrival (TOA) monitor for ultrafast photoelectron diffraction and other pump-probe experiments. Scoby, et al. has recently demonstrated TOA measurements for 3.5 MeV, <10 pC electron beams with <50 fs resolution [8]. With convenient 90-degree crossing geometry, they could capture both the longitudinal and transverse propagation of the bunch field, which shows a great progress in THz beam diagnostics.

Another important advance in single-shot THz diagnostics was presented by J. van Tilborg at BNL. Tilborg described a novel EO configuration based on THz-induced optical sideband generation. Instead of using a short optical probe pulse, a long optical pulse with a narrow bandwidth was mixed with a THz pulse to be characterized in an EO crystal, which generates optical sidebands via difference/sum frequency mixing. Those newly generated sidebands, which can be easily measured with an optical spectrometer, directly provide the THz spectrum. This technique, however, can not directly provide the temporal THz waveform (necessary for complete bunch characterizations) because the THz phase information is not fully recovered. Furthermore, the sideband generation efficiency needs to be improved for broadband THz detection. Nonetheless, this method can detect, in principle, up to ~100 THz, far exceeding the traditionally limited optical bandwidth, and holds promise for characterizing sub-10 fs electron beams.

## ELECTRON BEAM CONTROL AND DYNAMICS

Novel methods of electron beam emittance exchange and emittance "partitioning" have recently been under investigation by several groups, with multiple goals in mind. The concept of emittance exchange was first proposed by Y. Orlov in 1991 [9], and in 2002 M. Cornacchia and P. Emma [10] proposed a practical way to achieve transverse to longitudinal emittance exchange with 2 dogleg bends and a transverse deflecting mode (TM110) RF cavity between them. This was later improved upon by K.-J. Kim and A. Sessler in 2006 [11]. The first experimental demonstration of transverse to longitudinal emittance exchange using this method was demonstrated at the Fermilab A0 photoinjector. In this Workshop J. Ruan presented the latest results from the A0 photoinjector which exchanges horizontal and longitudinal emittance. In this case, the initial emittance $(\varepsilon_x, \varepsilon_y, \varepsilon_z) = (3.7, 3.2, 16.2)$ μm is transformed to $(\varepsilon_x, \varepsilon_y, \varepsilon_z) = (13.9, 4.7, 7.7)$ μm.

J. Power presented plans underway at the Argonne Wakefield Accelerator to generate longitudinally ramped bunches by using a transverse mask to shape the beam in transverse (x,y) space followed by an emittance exchanger of the type described by Kim and Sessler to generate a bunch with a tailored longitudinal profile. One of the motivations for this set of experiments is to improve the transformer ratio for dielectric wakefield acceleration schemes. The transformer ratio is the ratio of maximum witness beam energy gain over drive beam energy loss, and for a bunch with a gaussian profile this ratio is 2. Simulations by Power show that transformer ratios as high as ~10 are attainable with longitudinally ramped profiles generated by the proposed scheme. Another proposed experiment at this facility is to use a mask with multiple slits upstream of an emittance exchanger, so that the emittance exchanger produces bunches with energy bands which will enhance the operation of a higher harmonic FEL.

P. Piot, et al. have already performed an experiment at the Fermilab A0 photoinjector of the type proposed by Power at Argonne, and these results were also presented at this Workshop. In this case, they have placed a multiple slit mask upstream of the emittance exchanger to produce a closely spaced (~1 ps) train of bunches of ~200 fs width. This technique, used in conjunction with a flat beam transform, can potentially be used to excite terahertz radiation from dielectric slab structures.

D. Xiang proposed a modified version of the standard emittance exchanger by replacing the transverse deflecting mode cavity between the doglegs with an undulator in which a $TEM_{10}$ mode laser is allowed to interact with the beam. In this case, the laser field is equivalent to the transverse RF field near the zero crossing of the laser field and only a small fraction of the beam particles near this zero crossing will be emittance exchanged. He shows this beam can be used to enhance the operation of a compact X-ray FEL. In addition, such a scheme could in principle be used in a storage ring FEL.

B. Carlston presented several ideas for phase space partitioning, motivated by plans to build a 50 keV (0.024 nm) XFEL (MaRIE) proposed at Los Alamos. A typical photoinjector at 0.5 nC generates emittances of $(\varepsilon_x, \varepsilon_y, \varepsilon_z) = (0.7, 0.7, 1.4)$ μm for a phase space "volume" of ~0.7 μm$^3$, whereas the needs of the MaRIE project are $(\varepsilon_x, \varepsilon_y, \varepsilon_z) = (0.15, 0.15, 100)$ μm for a phase space "volume" of ~2.3 μm$^3$. This leaves ample room for phase space manipulations to obtain the desired emittance partitioning. Carlston presented investigations and issues with generalized and more complex schemes of emittance exchange and flat beam transforms to accomplish the appropriate emittance partitioning in a realizable way.

A novel bunch compression and synchronization scheme to be tested at the UCLA Pegasus Laboratory was presented by J Moody. In this scheme it is proposed to compress and synchronize a 3.5 MeV, 100 fs electron beam from a photoinjector via IFEL interaction with THz radiation in an undulator. A train of up to 16 THz pulses at 1 μJ/pulse is produced from a 800 nm, 2 mJ, 35 fs IR laser by optical rectification in a lithium niobate crystal.

Simulations show a compression ratio of 7 and jitter reduction of a factor of ~10. The motivation for this experiment is to inject beam into ps scale structures with minimal time jitter.

Presentations on the University of Maryland Electron Ring (UMER) status and experiments were given by R. Kishek, S. Bernal, B. Beaudoin, T. Koeth, K. Fiuza (poster), and H. Zhang (see previous section). UMER is a low energy (10 keV) electron storage ring with a thermionic gun as injector, and because it is such low energy it provides an excellent laboratory for the study of beam dynamics in the space charge dominated regime. The machine is operated in the current range of 0.6 mA to 100 mA with a tune shift of ~1 at the low current end. After substantial injection and ring tuning, the beam can now circulate for up to 1000 turns. Ongoing studies of longitudinal dynamics include the study of bunch edge erosion, density wave evolution, and longitudinal focusing. R. Kishek and others are characterizing the beam dynamics in the transverse phase space and have measured the evolution of tune space over many turns. Both $1^{st}$ and $2^{nd}$ order (incoherent) transverse resonances have been measured. When the beam is injected into the ring, space charge forces and momentum spread cause it to spread out longitudinally and fill the ring, creating DC beam. B. Beaudoin is developing a longitudinal focusing system based on an inductive cell to form "barrier RF buckets" to contain the beam longitudinally and hence improve the AC beam lifetime. T. Koeth is developing an RF knockout technique to measure the DC beam component.

Finally, C. Chen presented calculations on a new beam state called adiabatic thermal beam equilibrium. In this state for high brightness beams, particle motion in phase space is more stable (less chaotic) than previously investigated equilibrium beam states (Kapchinskij-Vladimirskij [12], for example) offering the potential for more stable beam envelopes, stable emittances, reduced beam losses, and reduction of halo formation. C. Chen and H. Wei have performed the calculations for both solenoidal focusing channels and alternating-gradient focusing channels, and some initial experiments have been performed at UMER and Spring-8 [13]. This beam state has potential application in high brightness elliptic guns and in beam-matching applications, among others.

## ACKNOWLEDGMENTS

We wish to thank all the participants and presenters of Working Group 5 for making this an instructive and successful meeting. Much more detail on their work may be found in their individual contributions to this Proceedings, and we apologize if we have neglected to fully do justice to their work in this summary. We also wish to thank the conference organizers for their untiring efforts in making this a successful workshop.